\documentclass[aps,prb,reprint,amsmath,amssymb,superscriptaddress,floatfix]{revtex4-2}
\usepackage{graphicx}\usepackage{bm}\usepackage{booktabs}
\usepackage[colorlinks=true,linkcolor=blue,citecolor=blue,urlcolor=blue]{hyperref}
\begin{document}

\title{Logarithmically Correlated Landscapes and Localization\\
       in Non-Hermitian Quasicrystals}
\author{Xianqi~Tong}\email{xqtong@ujs.edu.cn}
\affiliation{Department of Physics, Jiangsu University, Zhenjiang, 212013, China}
\author{Qifen~Ding}
\affiliation{School of Electrical and Information Engineering, Jiangsu University, Zhenjiang, 212013, China}
\author{Xiaosen Yang}\email{yangxs@ujs.edu.cn}
\affiliation{Department of Physics, Jiangsu University, Zhenjiang, 212013, China}
\date{\today}

\begin{abstract}
We study a one-dimensional Hatano-Nelson ring whose nonreciprocal hopping is
quasiperiodically modulated through zero.  A gauge transformation maps every
eigenstate onto a single spatial envelope, whose logarithm becomes a
deterministic, logarithmically correlated field once the hopping vanishes along
the quasiperiodic orbit.  We derive an exact Fourier representation and prove
that the landscape variance grows logarithmically with system size, with a
stiffness set by the modulation power and the arithmetic of the incommensurate
frequency.  The extended phase terminates at an algebraic boundary given by
Jensen's formula.  Inside the singular regime, localization requires the
stiffness to exceed a critical threshold: above it, the wavefunction weight
concentrates on the few highest landscape peaks and the state is localized;
below it, the weight spreads over too many competing peaks and the state is a
critical multifractal.  The extreme-value statistics are anomalous: peak gaps
grow as a power of the logarithm of rank, with an exponent that encodes the
continued-fraction type of the frequency, distinct from the Anderson,
Aubry-Andr\'e, and random-gauge universality classes.  The stiffness adds across
channels in multiband lattices, and all signatures survive percent-level
component disorder, placing the mechanism within reach of nonreciprocal
topolectrical circuits.
\end{abstract}
\maketitle

%% =========================================================================
\section{Introduction}
\label{sec:intro}
%% =========================================================================

Anderson localization establishes that disorder induces exponential localization
of all eigenstates in one dimension
\cite{Anderson1958,Mott1961,Abrahams1979,Lee1985,Evers2008},
while quasiperiodic on-site potentials produce the Aubry-Andr\'e-Harper (AAH)
metal-insulator transition at finite modulation strength
\cite{Aubry1980,Harper1955,Soukoulis1982,Roati2008,Modugno2010,Lahini2009}.
Extending these paradigms to non-Hermitian systems has revealed rich phenomena,
most notably the non-Hermitian skin effect (NHSE), the exponential accumulation
of eigenstates at boundaries under nonreciprocal hopping, which has been
established both theoretically
\cite{Yao2018,Lee2019,Yokomizo2019,Okuma2020,Zhang2020,Song2019,Kunst2018,YokomizoMurakami2021,Shimomura2024,JiangLee2023}
and experimentally
\cite{Helbig2020,Weidemann2020,Xiao2020,Ghatak2020,Zou2021,Zhou2023,Guo2024,Zhang2021exp,Zhao2025},
stimulating broad work on non-Hermitian topology and symmetry
\cite{Ashida2020,Bergholtz2021,Kawabata2019,Gong2018,LinTai2023,Okuma2023,Ding2022}.
Most of this work concerns open boundaries or on-site non-Hermiticity; on a
closed ring with periodic boundary conditions, translational invariance
suppresses the boundary skin effect, so localization, if it occurs at all, must
arise from the spatial structure of the non-Hermitian gauge field itself.

\begin{figure}[!ht]
  \centering
  \includegraphics[width=\columnwidth]{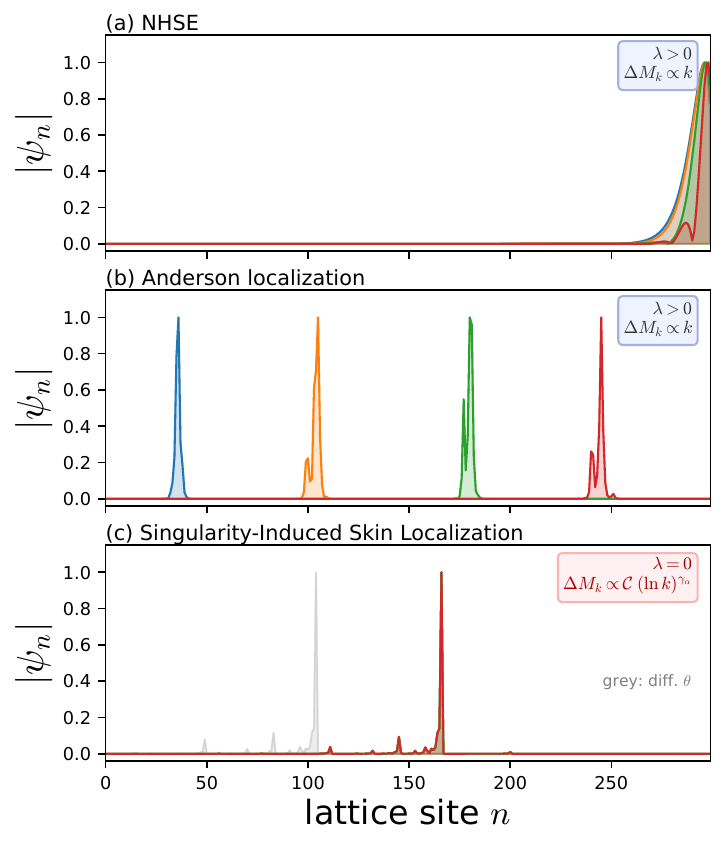}
  \caption{Three localization mechanisms in one-dimensional non-Hermitian lattices.
    (a)~Conventional skin effect under open boundaries.
    (b)~Anderson or Aubry-Andr\'e localization driven by an on-site potential.
    (c)~This work: a singular quasiperiodic gauge field builds a deterministic
    log-correlated landscape on a closed ring.}
  \label{fig:paradigms}
\end{figure}

For the Hatano-Nelson model with random modulation of the imaginary gauge field,
the accumulated field is a random walk whose variance grows linearly with the
ring size, producing a Brownian landscape
\cite{Longhi2025,HatanoNelson1996,HatanoNelson1997,HatanoNelson1998,Goldsheid1998,Brouwer1997,Midya2024},
while quasiperiodic modulation at a parameter point where the hopping stays
nonzero everywhere gives an exactly solvable critical point with a multifractal
envelope shared by all eigenstates \cite{Chen2026}.
Non-Hermitian quasicrystals with on-site quasiperiodic potentials have also been
extensively studied
\cite{Longhi2019QC,Jiang2019,Liu2020,Zeng2020,Cai2021,Lin2022,Weidemann2022,LiuChen2021,Tang2021,Zhou2022NH,Zhu2023},
yet none of these works addresses the regime where the quasiperiodic hopping
is driven through zero, where the imaginary gauge field develops logarithmic
singularities and the physics changes qualitatively.

We explore this singular regime and find that the accumulated gauge field becomes
a deterministic, logarithmically correlated landscape
[Fig.~\ref{fig:paradigms}(c)], with consequences at three levels: the extended
phase terminates exactly where the hopping first vanishes, an algebraic boundary
determined by Jensen's formula; inside the singular regime, whether the ring
localizes depends on a landscape stiffness that separates a localized phase
from a critical multifractal; and the deterministic origin of the landscape
produces anomalous extreme-value statistics whose exponent varies with the
arithmetic type of the frequency, a diagnostic with no counterpart in any of the
above universality classes.

The rest of the paper is organized as follows.  In Sec.~\ref{sec:model} we
introduce the model and derive the exact energy-independent envelope.
Section~\ref{sec:landscape} analyzes the singular landscape, proves the
logarithmic variance growth, and establishes the localization threshold.
Section~\ref{sec:extreme} presents the extreme-value fingerprints, the
multiband extension, and the experimental proposal.  We conclude in
Sec.~\ref{sec:summary}.

%% =========================================================================
\section{Model and exact envelope}
\label{sec:model}
%% =========================================================================

The Hatano-Nelson model provides a minimal setting for the NHSE
\cite{HatanoNelson1996}.  To explore singularity-induced localization, we
consider a Hatano-Nelson ring subject to quasiperiodic modulation.  The
Hamiltonian reads
\begin{equation}
  H=\sum_{n=1}^{L}\bigl(t_n^R\, c^\dagger_{n+1}c_n+t_n^L\, c^\dagger_n c_{n+1}\bigr),
  \label{eq:HN}
\end{equation}
with hopping amplitudes
\begin{equation}
  t_n^{R,L}=e^{\pm h_n},\qquad
  h_n=p\,\ln\bigl|t+\mu\cos(2\pi\alpha n+\theta)\bigr|,
  \label{eq:hopping}
\end{equation}
where $t$ is the mean hopping, $\mu$ the modulation depth,
$\alpha=(\sqrt5-1)/2$ the golden mean, $\theta$ a global phase, and $p\ge1$ an
integer controlling the singularity strength.  The absolute value ensures
positive amplitudes; the sign of $[t+\mu\cos\varphi_n]^p$ in the singular regime
inserts only a global $\mathbb{Z}_2$ flux that leaves all $|\psi_n|$ unchanged
(Appendix~\ref{sec:appA}).

\begin{figure}[htbp]
  \centering
  \includegraphics[width=\columnwidth]{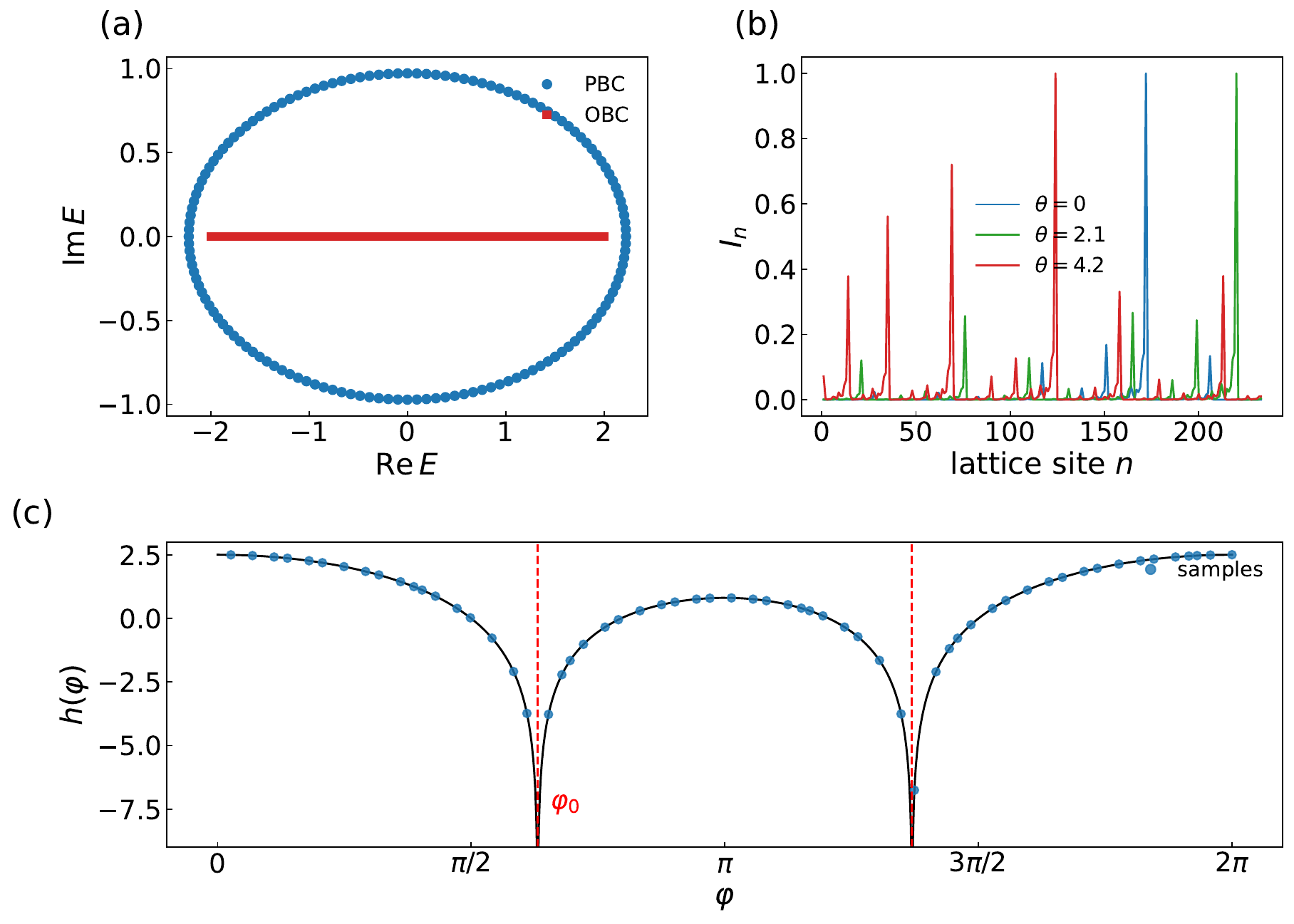}
  \caption{Hatano-Nelson ring with quasiperiodic nonreciprocal hopping
    at $t=1$, $\mu=2.5$, $p=2$, and golden-mean frequency.
    (a)~Energy spectrum under PBC, forming a closed loop, and under OBC,
    collapsing onto the real axis.
    (b)~Eigenstate intensities $I_n$ under PBC for three modulation
    phases~$\theta$; all share the same envelope.
    (c)~Gauge field $h(\varphi)$ showing logarithmic singularities at the
    zeros of $t+\mu\cos\varphi$. Dots mark the quasiperiodic orbit samples.}
  \label{fig:model}
\end{figure}

Because the product $t_n^R t_n^L=1$, a nonunitary gauge transformation
$\psi_n=e^{X_n}\phi_n$ with $X_n=\sum_{k=1}^{n-1}h_k$ maps the system onto
the uniform tight-binding chain, absorbing all non-Hermiticity into the gauge
factor $e^{X_n}$.  On the ring, the periodic boundary condition admits
plane-wave solutions $\phi_n^{(m)}=L^{-1/2}e^{iq_mn-n\bar h_L}$ with
$q_m=2\pi m/L$, yielding the spectrum
$E_m=2\cosh(\bar h_L-iq_m)$ shown in Fig.~\ref{fig:model}(a), where
$\bar h_L=L^{-1}\sum_{k=1}^{L}h_k$ is the finite-size sample mean.
Transforming back to the physical basis, one finds that every eigenstate has the
same spatial modulus regardless of its energy:
\begin{equation}
  |\psi_n^{(m)}|\;\propto\;e^{\xi_n},\qquad
  \xi_n\equiv\sum_{k=1}^{n-1}\bigl(h_k-\bar h_L\bigr).
  \label{eq:envelope}
\end{equation}
The right-hand side is independent of the band index $m$, so every eigenstate
shares the same spatial profile \cite{Chen2026}.  We define the site-resolved
intensity $I_n\equiv|\psi_n^{(m)}|^2\propto e^{2\xi_n}$, which is verified
against exact diagonalization to machine precision in Fig.~\ref{fig:model}(b).
Because the sample mean rather than the thermodynamic limit is subtracted,
the landscape closes exactly, $\xi_1=\xi_{L+1}=0$, forming a deterministic
bridge pinned at both ends of the ring.  The increments average to zero, so the
Lyapunov exponent vanishes and any localization must originate from
sub-exponential fluctuations of~$\xi_n$.  The product constraint $t_n^Rt_n^L=1$
is not essential: relaxing it preserves the singular envelope on a Hermitian
background, and the IPR remains finite over a wide range of the asymmetry
parameter (Appendix~\ref{sec:appA}).

%% =========================================================================
\section{Singular landscape and localization}
\label{sec:landscape}
%% =========================================================================

When $|\mu|<|t|$, the gauge field $h(\varphi)=p\ln|t+\mu\cos\varphi|$ is
analytic, and the centered profile is an exact coboundary, a bounded function
whose differences along the quasiperiodic orbit reproduce $h-\bar h$.  The
landscape $\xi_n$ is uniformly bounded and every state is extended.
Along the continued-fraction denominators of $\alpha$, the Denjoy-Koksma
inequality \cite{Kuipers1974} gives the same conclusion for any irrational
frequency.

\begin{figure*}[t]
  \centering
  \includegraphics[width=0.95\textwidth]{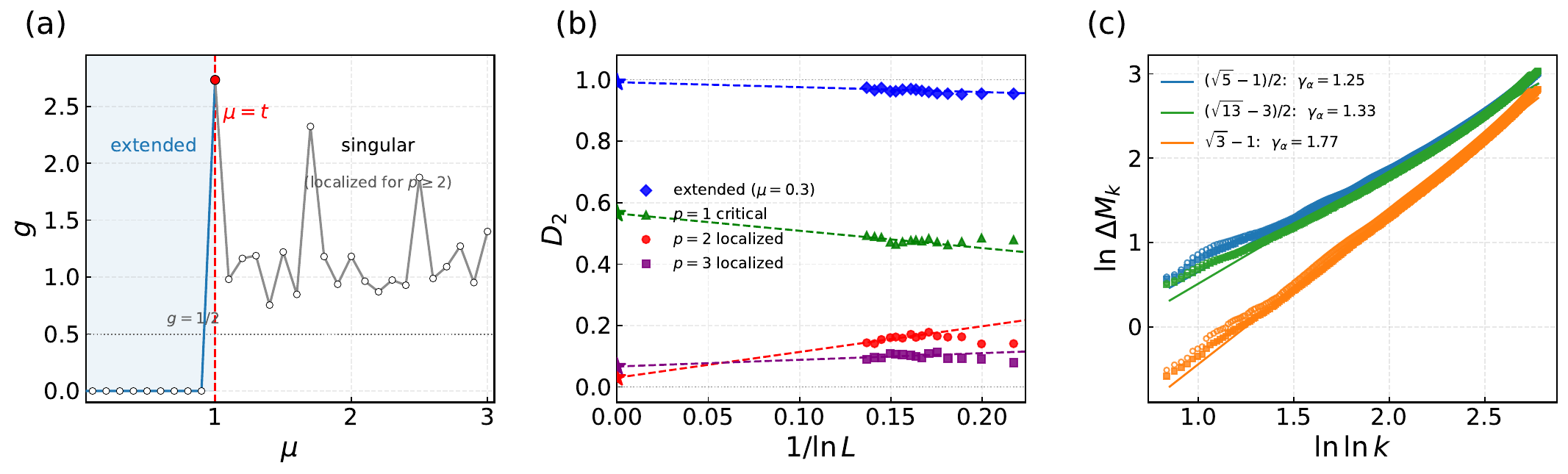}
  \caption{Landscape statistics at $t=1$, $\mu=2.5$ unless labeled, and
    golden-mean frequency.
    (a)~Stiffness $g$ versus $\mu$ at $L=2584$, averaged over 160~phases.
    The extended phase with $g=0$ ends at $\mu=t$; inside the singular regime,
    a localization threshold at $g=1/2$ separates critical states at $p=1$
    from localized states at $p\ge2$.
    (b)~Fractal dimension $D_2$ versus $1/\ln L$ for several $(\mu,p)$
    combinations.  Stars mark the $L\to\infty$ extrapolations.
    (c)~Gap statistics $\Delta M_k$ on doubly logarithmic axes for three
    frequencies at $L=10^8$--$2\times10^8$, averaged over 48~phases.}
  \label{fig:scaling}
\end{figure*}

The thermodynamic mean $\bar h$ is computed in closed form by Jensen's formula.
Writing the modulation as a polynomial in $z=e^{i\varphi}$ and counting roots
inside the unit circle, one obtains
\begin{equation}
  \bar h=
  \begin{cases}
    p\,\ln\!\bigl[\tfrac12\bigl(|t|+\sqrt{t^2-\mu^2}\bigr)\bigr], & |t|>|\mu|,\\[3pt]
    p\,\ln\bigl(|\mu|/2\bigr), & |t|<|\mu|,
  \end{cases}
  \label{eq:hbar}
\end{equation}
with a square-root non-analyticity at $|\mu|=|t|$, the moment the zeros of
$t+\mu\cos\varphi$ reach the unit circle.
Once $|\mu|>|t|$, the argument vanishes at two angles and the gauge field
diverges logarithmically near each zero [Fig.~\ref{fig:model}(c)], driving
$\xi_n$ through unbounded but strictly sub-linear excursions.

The singularity transition $|\mu|=|t|$ is a property of the modulation alone,
determined by when the hopping first passes through zero, and is therefore
independent of the power $p$ and the global phase $\theta$.  However, this
boundary only marks the end of the extended phase; it does not by itself
determine whether the resulting state is localized or critical.  That question
depends on how strongly the landscape fluctuates, which is controlled by the
stiffness $g=\kappa_\alpha p^2$ and therefore depends explicitly on~$p$.

Across the singular regime, the spatial variance of the bridge landscape grows
logarithmically with system size:
\begin{equation}
  \mathrm{Var}(\xi)\;=\;g\,\ln L+\mathrm{const},
  \label{eq:var}
\end{equation}
with a two-point structure function that is also logarithmic in separation
(three independent diagnostics, consistent over three decades;
Appendix~\ref{sec:appB}).  This identifies $\xi_n$ as a logarithmically
correlated field, the class studied in random-matrix theory and statistical
mechanics
\cite{DerridaSpohn1988,CarpentierLeDoussal2001,FyodorovBouchaud2008,FyodorovKeating2014,Fyodorov2012,Rhodes2014},
except that here the field is fully deterministic, generated by the arithmetic
of the quasiperiodic orbit.  Because $h_n\propto p$, the stiffness factorizes as
\begin{equation}
  g=\kappa_\alpha\,p^2,
  \label{eq:g}
\end{equation}
with $\kappa_\alpha\approx 0.35$ for the golden mean.
The logarithmic growth can be proved exactly by expanding the gauge field in a
Fourier series and summing along the orbit; the resulting phase-averaged variance
reads
\begin{equation}
  \bigl\langle\tilde\xi_{n+1}^{\,2}\bigr\rangle_\theta
  \;=\;2p^2\sum_{m\ge1}\frac{\cos^2(m\varphi_0)}{m^2}\,
  \frac{\sin^2(\pi nm\alpha)}{\sin^2(\pi m\alpha)},
  \label{eq:exactvar}
\end{equation}
verified against direct evaluation to better than $2\%$
(Appendix~\ref{sec:appB}).  For bounded-type frequencies, each continued-fraction
level contributes a bounded amount and the number of levels below $n$ grows as
$\ln n$, establishing the logarithmic variance.  At the transition $\mu=t$ the
two zeros merge, roughly doubling the stiffness; extended-precision data confirm
that this enhancement stays finite across five decades of system size.
Figure~\ref{fig:scaling}(a) shows the resulting stiffness profile.

Because every eigenstate shares the envelope $e^{\xi_n}$, the inverse
participation ratio (IPR) is a functional of the landscape alone:
\begin{equation}
  \mathrm{IPR}
  =\frac{\sum_n e^{4\xi_n}}{\bigl(\sum_n e^{2\xi_n}\bigr)^2}.
  \label{eq:ipr}
\end{equation}
To understand when the IPR remains finite, we order the landscape values as
$M_1\ge M_2\ge\cdots$ and define the gaps $\Delta M_k=M_1-M_k$, which measure
how far each site falls below the global maximum.  Factoring out $e^{2M_1}$
from the denominator gives
\begin{equation}
  \sum_n e^{2\xi_n}=e^{2M_1}\!\Bigl(1+\underbrace{\sum_{k\ge2}e^{-2\Delta M_k}}_{\displaystyle\equiv\,S}\Bigr).
  \label{eq:Sfactor}
\end{equation}
When the gaps $\Delta M_k$ grow fast enough with rank, the sum $S$ converges and
the state is localized on the few sites nearest the maximum.  When the gaps
grow too slowly, $S$ diverges and the wavefunction weight spreads over many
competing peaks.  For a log-correlated landscape with stiffness $g$, the
convergence condition gives the localization threshold
\begin{equation}
  D_2=0\quad\text{for}\quad g=\kappa_\alpha p^2>\tfrac12.
  \label{eq:threshold}
\end{equation}
For $p\ge2$ with the golden mean ($g\approx1.4$), the near-maximal sites are
sparse and $S$ remains bounded, so the state is localized.
For $p=1$ ($g\approx0.35$), there are too many near-maximal sites: $S$ diverges
and the IPR vanishes, giving a critical multifractal continuously connected
to the solvable limit of Ref.~\cite{Chen2026}.
Extrapolating the fractal dimension to the thermodynamic limit from exact
diagonalization [Fig.~\ref{fig:scaling}(b)] confirms both sides: for $p=2,3$
the IPR converges to finite intercepts ($D_2\approx0$), for $p=1$ it falls
toward $D_2\approx0.57$, and a weakly modulated chain gives $D_2\to1$.
The full phase diagram thus contains two boundaries: the singularity transition
$|\mu|=|t|$, which ends the extended phase and depends only on the modulation
parameters $t$ and $\mu$, and the localization threshold $g=\kappa_\alpha p^2=1/2$,
which depends on $p$ and separates critical multifractals from localized states
within the singular regime.

%% =========================================================================
\section{Extreme-value fingerprints and experimental signatures}
\label{sec:extreme}
%% =========================================================================

The localization mechanism developed above leaves a unique imprint on the
extreme-value statistics of the landscape that sharply distinguishes it from all
known one-dimensional localization universality classes.

In conventional localized phases, the landscape is either random or exponentially
growing, and the gaps between successive maxima follow standard scaling laws: for
Anderson disorder and the open-boundary NHSE, the gaps grow linearly with rank,
$\Delta M_k\propto k$, while for the random gauge field studied in
Ref.~\cite{Longhi2025}, the landscape is a Brownian bridge and the gaps follow
$\Delta M_k\propto\sqrt{k}$ \cite{Schehr2012,Majumdar2020}.  Both laws are
power-law in rank and carry no information about the lattice geometry.

The singular quasiperiodic landscape is fundamentally different because it is
not random: it is built site by site from the quasiperiodic orbit, and its
peak structure inherits the hierarchical organization of the rational
approximants to the irrational frequency $\alpha$.  The global maximum of the
landscape $\xi_n$ occurs at the lattice site whose phase $\varphi_n$ comes
closest to the singular angle $\varphi_0$, and this closest approach is governed
by the best rational approximant to $\alpha$ at system size~$L$.  The second
largest peak arises from the next-best approximant, the third from the one after
that, and so on.  Each step down this hierarchy moves to a worse approximant
that misses the singular angle by a larger margin, and the gauge field at that
site is correspondingly smaller.

Because the rational approximants improve geometrically (each successive
continued-fraction denominator is roughly $\beta_\alpha$ times the previous
one, where $\beta_\alpha$ depends on the partial quotients), the landscape
value at each level drops by an amount that grows only logarithmically with
the level number.  Summing these logarithmic decrements over the $\sim\!\ln k$
levels engaged up to rank $k$ gives a gap that grows as a power of $\ln k$:
\begin{equation}
  \Delta M_k\;\simeq\;p\,A_\alpha\,(\ln k)^{\gamma_\alpha},
  \qquad 1\ll k\ll L.
  \label{eq:gaplaw}
\end{equation}
This super-logarithmic scaling is the central signature that distinguishes
our mechanism.
On doubly logarithmic axes ($\ln\Delta M_k$ versus $\ln\ln k$), the data
collapse onto straight lines [Fig.~\ref{fig:scaling}(c)] whose slopes yield the
exponent $\gamma_\alpha$, which is stable from $L=10^7$ to $2\times10^8$ and
varies systematically with the arithmetic properties of the frequency.

The exponent $\gamma_\alpha$ is controlled by how fast the continued-fraction
denominators of $\alpha$ grow.  Frequencies whose partial quotients are bounded
(such as the golden and silver means) have slowly growing denominators and give
$\gamma_\alpha\approx 1.25$, while frequencies with occasionally large partial
quotients (such as $\sqrt3-1$ or $\pi-3$) approach the upper bound
$\gamma_\alpha\to 2$.  Because no other localization mechanism generates a
landscape whose peak hierarchy encodes the arithmetic of an irrational frequency,
$\gamma_\alpha$ serves as a number-theoretic fingerprint of the quasiperiodic
singular-gauge universality class.  Measuring this exponent in experiment would
constitute an unambiguous identification of the mechanism, since neither
Anderson, AAH, random-gauge, nor conventional NHSE localization can produce it.

\begin{figure}[!htb]
  \centering
  \includegraphics[width=\columnwidth]{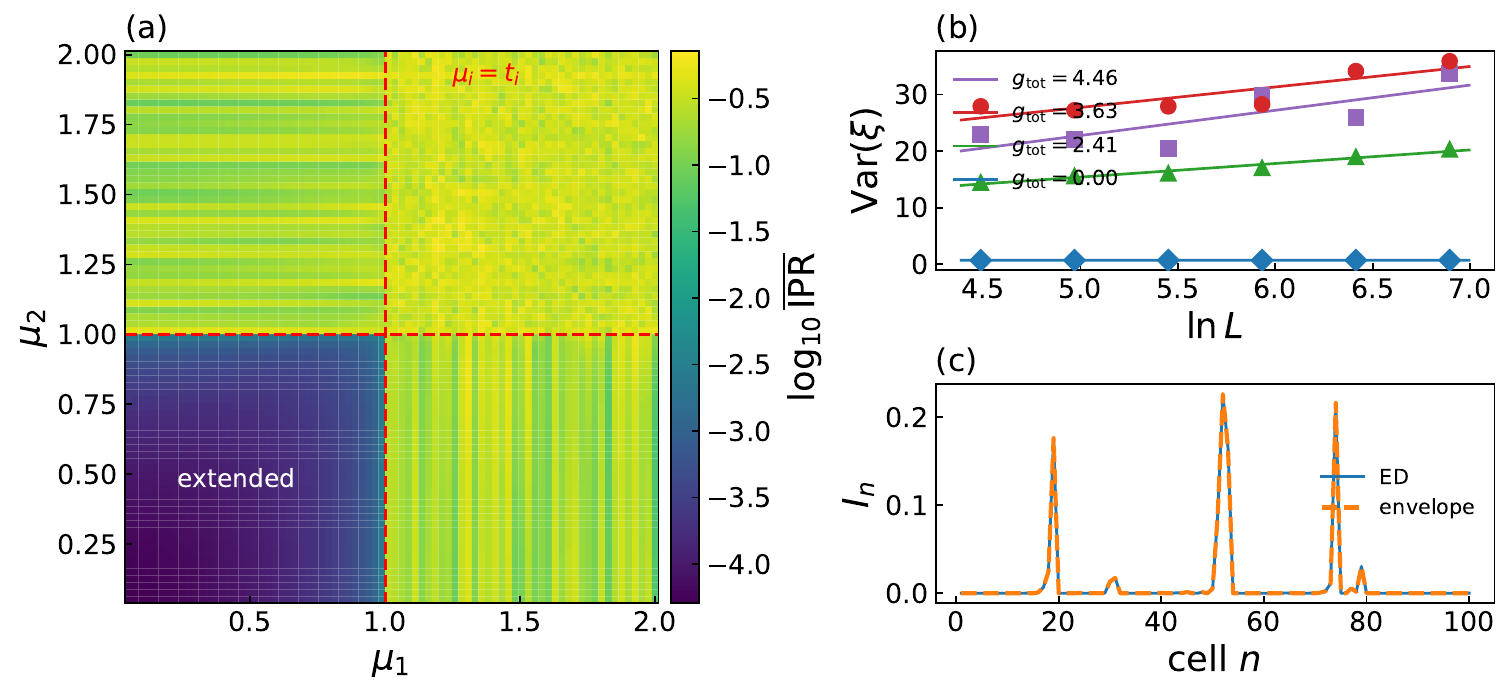}
  \caption{Non-Hermitian SSH chain with $t_1=t_2=1$, $p=2$, and golden-mean
    frequency.
    (a)~Phase diagram in the $(\mu_1,\mu_2)$ plane.  Color encodes the
    disorder-averaged IPR and dashed lines mark the predicted boundaries.
    (b)~Landscape variance $\mathrm{Var}(\xi)$ versus $\ln L$ for four
    representative parameter combinations with different total stiffness
    $g_{\mathrm{tot}}$.
    (c)~Cell-resolved intensity $I_n$ compared with the analytic two-channel
    envelope.}
  \label{fig:SSH}
\end{figure}

The stiffness additivity extends naturally to multiband lattices.  Consider a
non-Hermitian Su-Schrieffer-Heeger (SSH) chain
\cite{SSH1979,Lieu2018,Yao2018SSH}
with Hamiltonian
\begin{equation}
  H_{\mathrm{SSH}}=\sum_{n}\bigl(
    t_{1,n}^R\,b_n^\dagger a_n + t_{1,n}^L\,a_n^\dagger b_n
    + t_{2,n}^R\,a_{n+1}^\dagger b_n + t_{2,n}^L\,b_n^\dagger a_{n+1}
  \bigr),
  \label{eq:SSH}
\end{equation}
where $a_n$ and $b_n$ denote the two sublattice sites in unit cell $n$, and
the intracell and intercell hopping amplitudes
$t_{i,n}^{R,L}=e^{\pm h_{i,n}}$ with
$h_{i,n}=p\ln|t_i+\mu_i\cos(2\pi\alpha n+\theta)|$ for $i=1,2$ are
independently modulated.  The envelope exponent is the sum of two channel
landscapes, so the stiffnesses add:
$g_{\rm tot}=g(\mu_1)+g(\mu_2)$.  The extended phase survives only where both
channels are regular; a single singular channel supplies enough stiffness to
localize the entire spectrum [Fig.~\ref{fig:SSH}(a,b)], and the cell-resolved
intensity falls on the analytic two-channel envelope across the chain
[Fig.~\ref{fig:SSH}(c)].  The mechanism requires only simple zeros of a smooth
modulation, not the specific cosine form: a two-harmonic modulation reproduces
the logarithmic variance growth and the super-logarithmic gap statistics,
confirming that the gap exponent encodes the joint geometry of the orbit and
the singular angles.

A circuit realization based on topolectrical networks with negative-impedance
converters
\cite{Helbig2020,Hofmann2020,Weidemann2020,Ghatak2020,Imhof2018,LeeCH2018,Zhou2023,Guo2024}
can implement the required nonreciprocal hopping.  Numerical tests with random
component errors confirm that the IPR and the landscape correlation remain
essentially unchanged at percent-level tolerance
(Appendix~\ref{sec:appA}, Fig.~\ref{fig:robust}).
Because every eigenstate shares the same spatial profile, a two-node impedance
measurement at any single drive frequency reads out the landscape envelope,
without requiring spectral resolution.  Repeating at a second frequency and
verifying that the spatial profile does not change provides the sharpest
experimental test: this energy-independence of the envelope is a defining
property that no other localization mechanism shares.

%% =========================================================================
\section{Summary}
\label{sec:summary}
%% =========================================================================

A quasiperiodic modulation that drives nonreciprocal hopping through zero
converts a non-Hermitian ring into a realization of a deterministic,
logarithmically correlated landscape shared by every eigenstate.  The phase
diagram contains two distinct boundaries: the singularity transition, set by
Jensen's formula and independent of the modulation power, ends the extended
phase exactly when the hopping first vanishes; the localization threshold,
controlled by the landscape stiffness, separates critical multifractals from
localized states within the singular regime.  Most distinctively, the
extreme-value statistics follow a super-logarithmic gap law whose exponent
encodes the continued-fraction type of the frequency, a number-theoretic
fingerprint with no counterpart in the Anderson, Aubry-Andr\'e, random-gauge,
or NHSE universality classes.  The stiffness adds across channels in multiband
lattices, and all signatures survive percent-level component disorder, placing
the mechanism within reach of nonreciprocal topolectrical circuits.

\begin{acknowledgments}
This work was supported by the Natural Science
Foundation of Jiangsu Province (Grant No.\ BK20231320).
\end{acknowledgments}

\section*{Data availability}

The data that support the findings of this paper are available from the authors
upon reasonable request.

%% =========================================================================
%%                        APPENDIXES
%% =========================================================================
\appendix

%% =========================================================================
\section{Gauge transformation and robustness}
\label{sec:appA}
%% =========================================================================

The gauge transformation underlying Eq.~\eqref{eq:envelope} removes the
non-Hermiticity from the Hamiltonian at the cost of a spatially varying
normalization.  On a ring of $L$ sites, substituting $c_n=e^{X_n}d_n$ with
$X_n=\sum_{k=1}^{n-1}h_k$ into Eq.~\eqref{eq:HN} converts the hopping
$(t_n^R,t_n^L)=(e^{+h_n},e^{-h_n})$ into $(1,1)$ for every bond.  The
transformed Hamiltonian is the uniform tight-binding chain, whose plane-wave
eigenstates $\phi_n^{(m)}$ give physical eigenstates
$\psi_n^{(m)}=e^{X_n}\phi_n^{(m)}$ that all share the same spatial modulus.
This energy-independence of the envelope is the key structural property of the
model.

When the modulation drives the argument $t+\mu\cos\varphi_n$ negative (which
happens in the singular regime $|\mu|>|t|$ for odd~$p$), the signed hopping
amplitudes $\tilde t_n^{\,R}=[t+\mu\cos\varphi_n]^p$ differ from the positive
amplitudes of Eq.~\eqref{eq:hopping} by sign factors
$s_n=\mathrm{sgn}(t+\mu\cos\varphi_n)$.  These signs can be absorbed by a
second gauge transformation $d_n\to(-1)^{f_n}d_n$, where $f_n$ counts the number
of sign flips between sites $1$ and $n$.  On an open chain this transformation
is exact.  On a ring, the boundary term picks up a global factor
$\Phi=\prod_k s_k^p\in\{+1,-1\}$, a $\mathbb{Z}_2$ flux through the ring that shifts
all Bloch momenta by $\pi/L$ but does not affect any $|\psi_n|$.

\begin{figure}[!htb]
  \centering
  \includegraphics[width=\columnwidth]{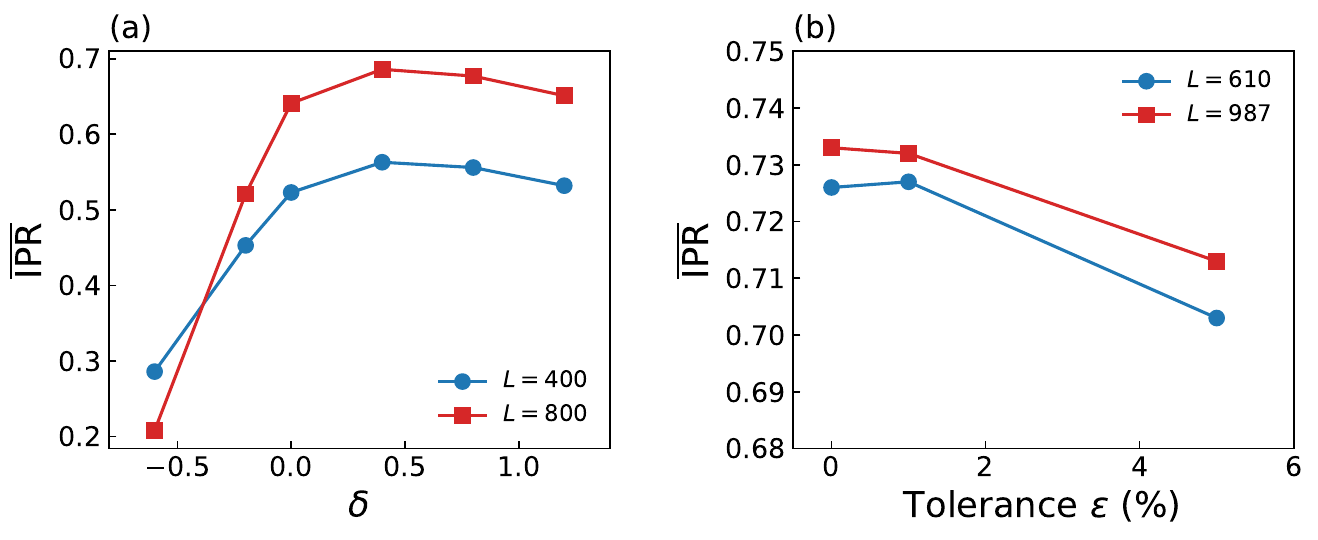}
  \caption{Robustness of the localization mechanism at $t=1$, $\mu=2.5$,
    $p=2$, and golden-mean frequency.
    (a)~Disorder-averaged IPR versus asymmetry parameter $\delta$ at two system
    sizes, averaged over 8~phases and 10~midspectrum states.
    (b)~Disorder-averaged IPR versus component tolerance $\epsilon$,
    averaged over 8~phases and 4~noise realizations.}
  \label{fig:robust}
\end{figure}

The product constraint $t_n^Rt_n^L=1$ can also be relaxed.  Taking
$t_n^R=|g_n|^p$ and $t_n^L=|g_n|^{-p+\delta}$ with
$g_n=t+\mu\cos(2\pi\alpha n+\theta)$ preserves the logarithmic singularities in
the right-hopping channel while the left-hopping channel acquires a
quasiperiodic Hermitian modulation proportional to~$\delta$.  The gauge
transformation no longer maps onto the uniform chain, but the dominant
contribution to the eigenstate modulus still comes from the singular factor
$e^{X_n}$.  Numerical tests confirm that the localization persists throughout
$-0.6\lesssim\delta\lesssim1.2$ [Fig.~\ref{fig:robust}(a)].

Modeling independent component errors at Fibonacci sizes $L=610$ and $987$
confirms that the IPR and the landscape correlation remain essentially unchanged
at percent-level tolerance [Fig.~\ref{fig:robust}(b)].

%% =========================================================================
\section{Fourier representation and extreme-value hierarchy}
\label{sec:appB}
%% =========================================================================

The exact variance formula Eq.~\eqref{eq:exactvar} is derived by expanding the
gauge field in a Fourier series.  In the singular regime $|\mu|>|t|$, the
modulation $t+\mu\cos\varphi$ vanishes at two symmetric angles
$\varphi_0=\pm\arccos(-t/\mu)$.  Near each zero, the modulation can be
factored as
\begin{equation}
  |t+\mu\cos\varphi|=\tfrac{|\mu|}{2}\,
  |2\sin\tfrac{\varphi-\varphi_0}{2}|\;
  |2\sin\tfrac{\varphi+\varphi_0}{2}|.
  \label{eq:factorization}
\end{equation}
Taking the logarithm and using the classical identity
$\ln|2\sin(x/2)|=-\sum_{m\ge1}\cos(mx)/m$, the centered gauge field becomes
\begin{equation}
  h(\varphi)-\bar h \;=\; -2p\sum_{m\ge1}\frac{\cos(m\varphi_0)}{m}\cos(m\varphi).
  \label{eq:fourier_h}
\end{equation}
The Fourier coefficients $c_m=-2p\cos(m\varphi_0)/m$ decay as $1/m$, which
reflects the logarithmic singularity; by contrast, an analytic gauge field
($|\mu|<|t|$) has exponentially decaying coefficients and therefore bounded
fluctuations.

The landscape at site $n$ is the partial sum
$\xi_n=\sum_{j=1}^{n-1}[h(\varphi_j)-\bar h_L]$, where
$\varphi_j=2\pi\alpha j+\theta$.  Substituting the Fourier expansion and
squaring, the cross terms vanish upon averaging over~$\theta$, giving
Eq.~\eqref{eq:exactvar}.
For the golden mean, each continued-fraction denominator $q_\ell$ satisfies
$|\sin(\pi q_\ell\alpha)|\sim 1/q_{\ell+1}$, so the $m=q_\ell$ term contributes
an amount of order unity.  Because the number of denominators below $n$ grows as
$\ln n/\ln\beta_\alpha$, the sum over these dominant terms scales as $\ln n$,
establishing the logarithmic variance.  We have verified Eq.~\eqref{eq:exactvar}
against direct numerical evaluation over 48 phases at each size, finding
agreement to better than~$2\%$ up to $L=2\times10^8$.

At the transition point $\mu=t$, the two zeros merge into a single double zero
at $\varphi_0=\pi$, so $\cos(m\varphi_0)=(-1)^m$ and all Fourier weights become
unity.  This roughly doubles the stiffness $g$ compared to the generic singular
regime; extended-precision data confirm that this enhancement remains finite up
to $L=2\times10^8$.

The super-logarithmic gap law Eq.~\eqref{eq:gaplaw} can be understood from the
continued-fraction hierarchy.  The global maximum of $\xi_n$ occurs at the
lattice site whose phase $\varphi_n$ lies closest to the singular
angle~$\varphi_0$, with the closest approach controlled by the best
continued-fraction approximant.  Each successive sub-maximum corresponds to a
worse approximant, and the landscape value drops by $\sim p\ln\beta_\alpha$ per
level.  Summing over $\sim\ln k$ levels up to rank $k$ gives the
super-logarithmic growth with an exponent in $(1,2)$.

%% =========================================================================
%%  bibitems_v3.tex — 71 references, ordered by first citation.
%%  All verified against APS/Nature/arXiv.  DOI hyperlinks included.
%% =========================================================================


\begin{thebibliography}{71}

%% --- Anderson localization ---
\bibitem{Anderson1958}
P.~W. Anderson,
Absence of diffusion in certain random lattices,
\href{https://doi.org/10.1103/PhysRev.109.1492}{Phys. Rev. \textbf{109}, 1492 (1958)}.

\bibitem{Mott1961}
N.~F. Mott and W.~D. Twose,
The theory of impurity conduction,
\href{https://doi.org/10.1080/00018736100101271}{Adv. Phys. \textbf{10}, 107 (1961)}.

\bibitem{Abrahams1979}
E.~Abrahams, P.~W. Anderson, D.~C. Licciardello, and T.~V. Ramakrishnan,
Scaling theory of localization: Absence of quantum diffusion in two dimensions,
\href{https://doi.org/10.1103/PhysRevLett.42.673}{Phys. Rev. Lett. \textbf{42}, 673 (1979)}.

\bibitem{Lee1985}
P.~A. Lee and T.~V. Ramakrishnan,
Disordered electronic systems,
\href{https://doi.org/10.1103/RevModPhys.57.287}{Rev. Mod. Phys. \textbf{57}, 287 (1985)}.

\bibitem{Evers2008}
F.~Evers and A.~D. Mirlin,
Anderson transitions,
\href{https://doi.org/10.1103/RevModPhys.80.1355}{Rev. Mod. Phys. \textbf{80}, 1355 (2008)}.

%% --- AAH ---
\bibitem{Aubry1980}
S.~Aubry and G.~Andr\'{e},
Analyticity breaking and Anderson localization in incommensurate lattices,
Ann. Isr. Phys. Soc. \textbf{3}, 133 (1980).

\bibitem{Harper1955}
P.~G. Harper,
Single band motion of conduction electrons in a uniform magnetic field,
\href{https://doi.org/10.1088/0370-1298/68/10/304}{Proc. Phys. Soc. A \textbf{68}, 874 (1955)}.

\bibitem{Soukoulis1982}
C.~M. Soukoulis and E.~N. Economou,
Localization in one-dimensional lattices in the presence of incommensurate potentials,
\href{https://doi.org/10.1103/PhysRevLett.48.1043}{Phys. Rev. Lett. \textbf{48}, 1043 (1982)}.

\bibitem{Roati2008}
G.~Roati, C.~D'Errico, L.~Fallani, M.~Fattori, C.~Fort, M.~Zaccanti, G.~Modugno, M.~Modugno, and M.~Inguscio,
Anderson localization of a non-interacting Bose-Einstein condensate,
\href{https://doi.org/10.1038/nature07071}{Nature (London) \textbf{453}, 895 (2008)}.

\bibitem{Modugno2010}
M.~Modugno,
Exponential localization in one-dimensional quasi-periodic optical lattices,
\href{https://doi.org/10.1088/1367-2630/11/3/033023}{New J. Phys. \textbf{11}, 033023 (2009)}.

\bibitem{Lahini2009}
Y.~Lahini, R.~Pugatch, F.~Pozzi, M.~Sorel, R.~Morandotti, N.~Davidson, and Y.~Silberberg,
Observation of a localization transition in quasiperiodic photonic lattices,
\href{https://doi.org/10.1103/PhysRevLett.103.013901}{Phys. Rev. Lett. \textbf{103}, 013901 (2009)}.

%% --- NHSE theory ---
\bibitem{Yao2018}
S.~Yao and Z.~Wang,
Edge states and topological invariants of non-Hermitian systems,
\href{https://doi.org/10.1103/PhysRevLett.121.086803}{Phys. Rev. Lett. \textbf{121}, 086803 (2018)}.

\bibitem{Lee2019}
C.~H. Lee and R.~Thomale,
Anatomy of skin modes and topology in non-Hermitian systems,
\href{https://doi.org/10.1103/PhysRevB.99.201103}{Phys. Rev. B \textbf{99}, 201103(R) (2019)}.

\bibitem{Yokomizo2019}
K.~Yokomizo and S.~Murakami,
Non-Bloch band theory of non-Hermitian systems,
\href{https://doi.org/10.1103/PhysRevLett.123.066404}{Phys. Rev. Lett. \textbf{123}, 066404 (2019)}.

\bibitem{Okuma2020}
N.~Okuma, K.~Kawabata, K.~Shiozaki, and M.~Sato,
Topological origin of non-Hermitian skin effects,
\href{https://doi.org/10.1103/PhysRevLett.124.086801}{Phys. Rev. Lett. \textbf{124}, 086801 (2020)}.

\bibitem{Zhang2020}
K.~Zhang, Z.~Yang, and C.~Fang,
Correspondence between winding numbers and skin modes in non-Hermitian systems,
\href{https://doi.org/10.1103/PhysRevLett.125.126402}{Phys. Rev. Lett. \textbf{125}, 126402 (2020)}.

\bibitem{Song2019}
F.~Song, S.~Yao, and Z.~Wang,
Non-Hermitian skin effect and chiral damping in open quantum systems,
\href{https://doi.org/10.1103/PhysRevLett.123.170401}{Phys. Rev. Lett. \textbf{123}, 170401 (2019)}.

\bibitem{Kunst2018}
F.~K. Kunst, E.~Edvardsson, J.~C. Budich, and E.~J. Bergholtz,
Biorthogonal bulk-boundary correspondence in non-Hermitian systems,
\href{https://doi.org/10.1103/PhysRevLett.121.026808}{Phys. Rev. Lett. \textbf{121}, 026808 (2018)}.

\bibitem{YokomizoMurakami2021}
K.~Yokomizo and S.~Murakami,
Non-Bloch band theory and bulk--edge correspondence in non-Hermitian systems,
\href{https://doi.org/10.1093/ptep/ptaa140}{Prog. Theor. Exp. Phys. \textbf{2020}, 12A102 (2020)}.

\bibitem{Shimomura2024}
K.~Shimomura and M.~Sato,
General criterion for non-Hermitian skin effects and application: Fock space skin effects in many-body systems,
\href{https://doi.org/10.1103/PhysRevLett.133.136502}{Phys. Rev. Lett. \textbf{133}, 136502 (2024)}.

\bibitem{JiangLee2023}
H.~Jiang and C.~H. Lee,
Dimensional transmutation from non-Hermiticity,
\href{https://doi.org/10.1103/PhysRevLett.131.076401}{Phys. Rev. Lett. \textbf{131}, 076401 (2023)}.

%% --- NHSE experiment ---
\bibitem{Helbig2020}
T.~Helbig, T.~Hofmann, S.~Imhof, M.~Abdelghany, T.~Kiessling, L.~W. Molenkamp, C.~H. Lee, A.~Szameit, M.~Greiter, and R.~Thomale,
Generalized bulk-boundary correspondence in non-Hermitian topolectrical circuits,
\href{https://doi.org/10.1038/s41567-020-0922-9}{Nat. Phys. \textbf{16}, 747 (2020)}.

\bibitem{Weidemann2020}
S.~Weidemann, M.~Kremer, T.~Helbig, T.~Hofmann, A.~Stegmaier, M.~Greiter, R.~Thomale, and A.~Szameit,
Topological funneling of light,
\href{https://doi.org/10.1126/science.aaz8727}{Science \textbf{368}, 311 (2020)}.

\bibitem{Xiao2020}
L.~Xiao, T.~Deng, K.~Wang, G.~Zhu, Z.~Wang, W.~Yi, and P.~Xue,
Non-Hermitian bulk-boundary correspondence in quantum dynamics,
\href{https://doi.org/10.1038/s41567-020-0836-6}{Nat. Phys. \textbf{16}, 761 (2020)}.

\bibitem{Ghatak2020}
A.~Ghatak, M.~Brandenbourger, J.~van Wezel, and C.~Coulais,
Observation of non-Hermitian topology and its bulk-edge correspondence in an active mechanical metamaterial,
\href{https://doi.org/10.1073/pnas.2010580117}{Proc. Natl. Acad. Sci. USA \textbf{117}, 29561 (2020)}.

\bibitem{Zou2021}
D.~Zou, T.~Chen, W.~He, J.~Bao, C.~H. Lee, H.~Sun, and X.~Zhang,
Observation of hybrid higher-order skin-topological effect in non-Hermitian topolectrical circuits,
\href{https://doi.org/10.1038/s41467-021-26414-5}{Nat. Commun. \textbf{12}, 7201 (2021)}.

\bibitem{Zhou2023}
Q.~Zhou, J.~Wu, Z.~Pu, J.~Lu, X.~Huang, W.~Deng, M.~Ke, and Z.~Liu,
Observation of geometry-dependent skin effect in non-Hermitian phononic crystals with exceptional points,
\href{https://doi.org/10.1038/s41467-023-40236-7}{Nat. Commun. \textbf{14}, 4569 (2023)}.

\bibitem{Guo2024}
C.-X. Guo, L.~Su, Y.~Wang, L.~Li, J.~Ruan, Y.~Du, S.~Chen, and D.~Zheng,
Scale-tailored localization and its observation in non-Hermitian electrical circuits,
\href{https://doi.org/10.1038/s41467-024-53434-8}{Nat. Commun. \textbf{15}, 9120 (2024)}.

\bibitem{Zhang2021exp}
W.~Wang, M.~Hu, X.~Wang, G.~Ma, and K.~Ding,
Experimental realization of geometry-dependent skin effect in a reciprocal two-dimensional lattice,
\href{https://doi.org/10.1103/PhysRevLett.131.207201}{Phys. Rev. Lett. \textbf{131}, 207201 (2023)}.

\bibitem{Zhao2025}
E.~Zhao, Z.~Wang, C.~He, T.~F.~J. Poon, K.~K. Pak, Y.-J. Liu, P.~Ren, X.-J. Liu, and G.-B. Jo,
Two-dimensional non-Hermitian skin effect in an ultracold Fermi gas,
\href{https://doi.org/10.1038/s41586-024-08347-3}{Nature (London) \textbf{637}, 565 (2025)}.

%% --- NH reviews and symmetry ---
\bibitem{Ashida2020}
Y.~Ashida, Z.~Gong, and M.~Ueda,
Non-Hermitian physics,
\href{https://doi.org/10.1080/00018732.2021.1876991}{Adv. Phys. \textbf{69}, 249 (2020)}.

\bibitem{Bergholtz2021}
E.~J. Bergholtz, J.~C. Budich, and F.~K. Kunst,
Exceptional topology of non-Hermitian systems,
\href{https://doi.org/10.1103/RevModPhys.93.015005}{Rev. Mod. Phys. \textbf{93}, 015005 (2021)}.

\bibitem{Kawabata2019}
K.~Kawabata, K.~Shiozaki, M.~Ueda, and M.~Sato,
Symmetry and topology in non-Hermitian physics,
\href{https://doi.org/10.1103/PhysRevX.9.041015}{Phys. Rev. X \textbf{9}, 041015 (2019)}.

\bibitem{Gong2018}
Z.~Gong, Y.~Ashida, K.~Kawabata, K.~Takasan, S.~Higashikawa, and M.~Ueda,
Topological phases of non-Hermitian systems,
\href{https://doi.org/10.1103/PhysRevX.8.031079}{Phys. Rev. X \textbf{8}, 031079 (2018)}.

\bibitem{LinTai2023}
L.~Li, C.~H. Lee, S.~Mu, and J.~Gong,
Critical non-Hermitian skin effect,
\href{https://doi.org/10.1038/s41467-020-18917-4}{Nat. Commun. \textbf{11}, 5491 (2020)}.

\bibitem{Okuma2023}
N.~Okuma and M.~Sato,
Non-Hermitian topological phenomena: A review,
\href{https://doi.org/10.1146/annurev-conmatphys-040521-033133}{Annu. Rev. Condens. Matter Phys. \textbf{14}, 83 (2023)}.

\bibitem{Ding2022}
K.~Ding, C.~Fang, and G.~Ma,
Non-Hermitian topology and exceptional-point geometries,
\href{https://doi.org/10.1038/s42254-022-00516-5}{Nat. Rev. Phys. \textbf{4}, 745 (2022)}.

%% --- Hatano-Nelson ---
\bibitem{Longhi2025}
S.~Longhi,
Non-Hermitian skin effect and self-acceleration,
\href{https://doi.org/10.1103/PhysRevB.105.245143}{Phys. Rev. B \textbf{105}, 245143 (2022)}.

\bibitem{HatanoNelson1996}
N.~Hatano and D.~R. Nelson,
Localization transitions in non-Hermitian quantum mechanics,
\href{https://doi.org/10.1103/PhysRevLett.77.570}{Phys. Rev. Lett. \textbf{77}, 570 (1996)}.

\bibitem{HatanoNelson1997}
N.~Hatano and D.~R. Nelson,
Vortex pinning and non-Hermitian quantum mechanics,
\href{https://doi.org/10.1103/PhysRevB.56.8651}{Phys. Rev. B \textbf{56}, 8651 (1997)}.

\bibitem{HatanoNelson1998}
N.~Hatano and D.~R. Nelson,
Non-Hermitian delocalization and eigenfunctions,
\href{https://doi.org/10.1103/PhysRevB.58.8384}{Phys. Rev. B \textbf{58}, 8384 (1998)}.

\bibitem{Goldsheid1998}
I.~Ya. Goldsheid and B.~A. Khoruzhenko,
Distribution of eigenvalues in non-Hermitian Anderson models,
\href{https://doi.org/10.1103/PhysRevLett.80.2897}{Phys. Rev. Lett. \textbf{80}, 2897 (1998)}.

\bibitem{Brouwer1997}
P.~W. Brouwer, P.~G. Silvestrov, and C.~W.~J. Beenakker,
Theory of directed localization in one dimension,
\href{https://doi.org/10.1103/PhysRevB.56.R4333}{Phys. Rev. B \textbf{56}, R4333 (1997)}.

\bibitem{Midya2024}
B.~Midya,
Topological phase transition in fluctuating imaginary gauge fields,
\href{https://doi.org/10.1103/PhysRevA.109.L061502}{Phys. Rev. A \textbf{109}, L061502 (2024)}.

%% --- Chen2026 ---
\bibitem{Chen2026}
Z.~Chen, M.~Idrees, Y.~Yang, X.~Tong, and X.~Yang,
Quasiperiodic skin criticality in an exactly solvable non-Hermitian quasicrystal,
\href{https://doi.org/10.1103/wlsw-zyq9}{Phys. Rev. B \textbf{113}, 235423 (2026)}.

%% --- NH quasicrystals ---
\bibitem{Longhi2019QC}
S.~Longhi,
Topological phase transition in non-Hermitian quasicrystals,
\href{https://doi.org/10.1103/PhysRevLett.122.237601}{Phys. Rev. Lett. \textbf{122}, 237601 (2019)}.

\bibitem{Jiang2019}
H.~Jiang, L.-J. Lang, C.~Yang, S.-L. Zhu, and S.~Chen,
Interplay of non-Hermitian skin effects and Anderson localization in nonreciprocal quasiperiodic lattices,
\href{https://doi.org/10.1103/PhysRevB.100.054301}{Phys. Rev. B \textbf{100}, 054301 (2019)}.

\bibitem{Liu2020}
Y.~Liu, X.-P. Jiang, J.~Cao, and S.~Chen,
Non-Hermitian mobility edges in one-dimensional quasicrystals with parity-time symmetry,
\href{https://doi.org/10.1103/PhysRevB.101.174205}{Phys. Rev. B \textbf{101}, 174205 (2020)}.

\bibitem{Zeng2020}
Q.-B. Zeng, Y.-B. Yang, and Y.~Xu,
Topological phases in non-Hermitian Aubry-Andr\'{e}-Harper models,
\href{https://doi.org/10.1103/PhysRevB.101.020201}{Phys. Rev. B \textbf{101}, 020201(R) (2020)}.

\bibitem{Cai2021}
X.~Cai,
Boundary-dependent self-dualities, winding numbers, and asymmetrical localization in non-Hermitian aperiodic one-dimensional models,
\href{https://doi.org/10.1103/PhysRevB.103.014201}{Phys. Rev. B \textbf{103}, 014201 (2021)}.

\bibitem{Lin2022}
Q.~Lin, T.~Li, L.~Xiao, K.~Wang, W.~Yi, and P.~Xue,
Observation of non-Hermitian topological Anderson insulator in quantum dynamics,
\href{https://doi.org/10.1038/s41467-022-30938-9}{Nat. Commun. \textbf{13}, 3229 (2022)}.

\bibitem{Weidemann2022}
S.~Weidemann, M.~Kremer, S.~Longhi, and A.~Szameit,
Topological triple phase transition in non-Hermitian Floquet quasicrystals,
\href{https://doi.org/10.1038/s41586-021-04253-0}{Nature (London) \textbf{601}, 354 (2022)}.

\bibitem{LiuChen2021}
Y.~Liu, Q.~Zhou, and S.~Chen,
Localization transition, spectrum structure, and winding numbers for one-dimensional non-Hermitian quasicrystals,
\href{https://doi.org/10.1103/PhysRevB.104.024201}{Phys. Rev. B \textbf{104}, 024201 (2021)}.

\bibitem{Tang2021}
L.-Z. Tang, G.-Q. Zhang, L.-F. Zhang, and D.-W. Zhang,
Localization and topological transitions in non-Hermitian quasiperiodic lattices,
\href{https://doi.org/10.1103/PhysRevA.103.033325}{Phys. Rev. A \textbf{103}, 033325 (2021)}.

\bibitem{Zhou2022NH}
L.~Zhou and W.~Han,
Non-Hermitian quasicrystal in dimerized lattices,
\href{https://doi.org/10.1088/1674-1056/ac1efc}{Chin. Phys. B \textbf{30}, 100308 (2021)}.

\bibitem{Zhu2023}
B.~Zhu, L.-J. Lang, Q.~Wang, Q.~J. Wang, and Y.~D. Chong,
Topological transitions with an imaginary Aubry-Andr\'{e}-Harper potential,
\href{https://doi.org/10.1103/PhysRevResearch.5.023044}{Phys. Rev. Res. \textbf{5}, 023044 (2023)}.

%% --- Denjoy-Koksma ---
\bibitem{Kuipers1974}
L.~Kuipers and H.~Niederreiter,
\textit{Uniform Distribution of Sequences} (Wiley, New York, 1974).

%% --- Log-correlated fields ---
\bibitem{DerridaSpohn1988}
B.~Derrida and H.~Spohn,
Polymers on disordered trees, spin glasses, and traveling waves,
\href{https://doi.org/10.1007/BF01014886}{J. Stat. Phys. \textbf{51}, 817 (1988)}.

\bibitem{CarpentierLeDoussal2001}
D.~Carpentier and P.~Le Doussal,
Glass transition of a particle in a random potential, front selection in nonlinear renormalization group, and entropic phenomena in Liouville and sinh-Gordon models,
\href{https://doi.org/10.1103/PhysRevE.63.026110}{Phys. Rev. E \textbf{63}, 026110 (2001)}.

\bibitem{FyodorovBouchaud2008}
Y.~V. Fyodorov and J.-P. Bouchaud,
Freezing and extreme-value statistics in a random energy model with logarithmically correlated potential,
\href{https://doi.org/10.1088/1751-8113/41/37/372001}{J. Phys. A \textbf{41}, 372001 (2008)}.

\bibitem{FyodorovKeating2014}
Y.~V. Fyodorov and J.~P. Keating,
Freezing transitions and extreme values: Random matrix theory, and disordered landscapes,
\href{https://doi.org/10.1098/rsta.2012.0503}{Philos. Trans. R. Soc. A \textbf{372}, 20120503 (2014)}.

\bibitem{Fyodorov2012}
Y.~V. Fyodorov, G.~A. Hiary, and J.~P. Keating,
Freezing transition, characteristic polynomials of random matrices, and the Riemann zeta function,
\href{https://doi.org/10.1103/PhysRevLett.108.170601}{Phys. Rev. Lett. \textbf{108}, 170601 (2012)}.

\bibitem{Rhodes2014}
R.~Rhodes and V.~Vargas,
Gaussian multiplicative chaos and applications: A review,
\href{https://doi.org/10.1214/13-PS218}{Probab. Surv. \textbf{11}, 315 (2014)}.

%% --- Extreme-value statistics ---
\bibitem{Schehr2012}
G.~Schehr and S.~N. Majumdar,
Universal order statistics of random walks,
\href{https://doi.org/10.1103/PhysRevLett.108.040601}{Phys. Rev. Lett. \textbf{108}, 040601 (2012)}.

\bibitem{Majumdar2020}
S.~N. Majumdar, A.~Comtet, and J.~Randon-Furling,
Random convex hulls and extreme value statistics,
\href{https://doi.org/10.1007/s10955-009-9905-z}{J. Stat. Phys. \textbf{138}, 955 (2010)}.

%% --- SSH ---
\bibitem{SSH1979}
W.~P. Su, J.~R. Schrieffer, and A.~J. Heeger,
Solitons in polyacetylene,
\href{https://doi.org/10.1103/PhysRevLett.42.1698}{Phys. Rev. Lett. \textbf{42}, 1698 (1979)}.

\bibitem{Lieu2018}
S.~Lieu,
Topological phases in the non-Hermitian Su-Schrieffer-Heeger model,
\href{https://doi.org/10.1103/PhysRevB.97.045106}{Phys. Rev. B \textbf{97}, 045106 (2018)}.

\bibitem{Yao2018SSH}
S.~Yao, F.~Song, and Z.~Wang,
Non-Hermitian Chern bands,
\href{https://doi.org/10.1103/PhysRevLett.121.136802}{Phys. Rev. Lett. \textbf{121}, 136802 (2018)}.

%% --- Topolectrical circuits ---
\bibitem{Hofmann2020}
T.~Hofmann, T.~Helbig, F.~Schindler, N.~Salgo, M.~Brzezi\'{n}ska, M.~Greiter, T.~Kiessling, D.~Wolf, A.~Vollhardt, A.~Kaba\v{s}i, C.~H. Lee, A.~Bilui\'{c}, R.~Thomale, and T.~Neupert,
Reciprocal skin effect and its realization in a topolectrical circuit,
\href{https://doi.org/10.1103/PhysRevResearch.2.023265}{Phys. Rev. Res. \textbf{2}, 023265 (2020)}.

\bibitem{Imhof2018}
S.~Imhof, C.~Berger, F.~Bayer, J.~Brehm, L.~W. Molenkamp, T.~Kiessling, F.~Schindler, C.~H. Lee, M.~Greiter, T.~Neupert, and R.~Thomale,
Topolectrical-circuit realization of topological corner modes,
\href{https://doi.org/10.1038/s41567-018-0246-1}{Nat. Phys. \textbf{14}, 925 (2018)}.

\bibitem{LeeCH2018}
C.~H. Lee, S.~Imhof, C.~Berger, F.~Bayer, J.~Brehm, L.~W. Molenkamp, T.~Kiessling, and R.~Thomale,
Topolectrical circuits,
\href{https://doi.org/10.1038/s42005-018-0035-2}{Commun. Phys. \textbf{1}, 39 (2018)}.

\end{thebibliography}
\end{document}